# Un discours et un public « Gilets Jaunes » au cœur du Grand Débat National ?

# Combinaison des approches IA et textométriques pour l'analyse de discours des plateformes « Grand Débat National » et « Vrai débat »

Mathieu Brugidou, Philippe Suignard, Caroline Escoffier et Lou Charaudeau

EDF R&D - Boulevard Gaspard Monge 91 120 Palaiseau – France

mathieu.brugidou@edf.fr, lou.charaudeau@edf.fr, caroline.escoffier@edf.fr, philippe.suignard@edf.fr

## Abstract 1 (in English)

In this contribution, we propose to analyze the statements coming from two "civic tech" platforms - the governmental platform, "Grand Débat National" and, its political and algorithmic response proposed by a Yellow Vest collective, "Vrai Débat" -, by confronting two families of algorithms dedicated to text analysis. We propose to implement, on the one hand, proven approaches in textual data analysis (Reinert/Iramuteq Method) which have recently shown their interest in the analysis of very large corpora and, on the other hand, new methods resulting from the crossroads of the computer worlds, artificial intelligence and automatic language processing. We will examine the methodological solutions for qualifying the social properties of speakers about whom we have little direct information. Finally, we will attempt to present some research questions at the crossroads of the political sociology of public opinion and data science, which such a confrontation opens up.

## Résumé

Dans cette contribution, nous nous proposons d'analyser les énoncés issus de deux plateformes de « civic tech » - plateforme gouvernementale, le « Grand Débat National » et, sa riposte politique et algorithmique proposée par un collectif de Gilets Jaunes, le « Vrai Débat »-, en confrontant deux familles d'algorithmes dédiées à l'analyse de textes. Nous nous proposons de mettre en œuvre, d'une part, des approches éprouvées en analyses des données textuelles (Méthode Reinert sous Iramuteq) qui ont montré récemment leur intérêt pour l'analyse de très grand corpus et, d'autre part, des méthodes nouvelles issues du croisement des mondes informatiques, de l'intelligence artificielle et du traitement automatique des langues. Nous nous interrogerons en outre sur les solutions méthodologiques permettant de qualifier les propriétés sociales des locuteurs sur lesquelles nous n'avons que peu d'information directe. Enfin, nous tenterons de présenter quelques questions de recherche au croisement de la sociologie politique de l'opinion publique mais aussi des data science, qu'une telle confrontation ouvre.

**Mots clés :** Machine Learning, Analyse des données textuelles, Civic Tech, Débat public, Gilets Jaunes.

# 1. Introduction

Dans cet article, nous nous proposons d'analyser les énoncés issus de deux plateformes de « civic tech » (Benvegnu, 2011 ; Mabi, 2014) – la plateforme gouvernementale du « Grand Débat National » (GDN) et sa riposte politique et algorithmique proposée par un collectif de Gilets Jaunes (GJ), le « Vrai Débat » (VD), en combinant deux familles d'algorithmes dédiées à l'analyse de textes. Nous nous proposons de mettre en œuvre, d'une part, des approches éprouvées en analyse des données textuelles (Méthode Reinert sous Iramuteq) qui ont montré récemment leur intérêt pour l'analyse de très grand corpus (Sebbah *et al.*, 2019 ; Brugidou, 2011) et, d'autre part, des méthodes nouvelles issues du croisement des mondes informatiques, de l'intelligence artificielle et du traitement automatique des langues (Cointet et Parasie, 2018).

Nous nous interrogerons notamment sur les solutions méthodologiques permettant de qualifier les propriétés sociales des locuteurs sur lesquelles nous n'avons que peu d'information directe – ce qui constitue pour l'analyse sociologique de ces plateformes mais aussi pour l'analyse des discours recueillis sur le web un obstacle de taille.  A partir, du corpus « Entendre la France », issu d'une troisième plateforme proposant un design identique à celui du Grand Débat National, nous disposons d'informations sur les propriétés sociales (âge, sexe notamment) et politiques (soutien aux Gilets Jaunes). Grâce à des algorithmes d'apprentissage, il est possible d'attribuer ces propriétés comme autant de probabilités aux participants du Grand Débat National. L'analyse de ces propriétés réaffectées comme variables illustratives (mots étoilés) au corpus du GDN devrait nous permettre d'avoir une première appréciation sur la pertinence de ces stratégies. Nous chercherons ainsi à préciser la présence spéculaire d'un discours mais aussi d'un public Gilets Jaunes au cœur même du dispositif GDN.

# 2. Hypothèses et corpus

Pour tenter de sortir de la crise politique suscitée par le mouvement des « Gilets Jaunes », le Président de la République a appelé à un « Grand Débat National » comportant notamment des réunions publiques mais aussi une importante phase de débat numérique à travers la mise en place d'une plateforme de délibération en ligne (le « Grand Débat »[1]). Des représentants des « Gilets Jaunes » ont par ailleurs répliqué en proposant une plateforme de débat alternative intitulée le « Vrai Débat »[2]. Très rapidement, et notamment devant le succès de ces plateformes et l'importance des corpus recueillis (cf. plus bas les corpus), la question de l'analyse et de la synthèse des propositions ou des échanges recueillis sur ces plateformes s'est posée : le recours aux méthodes issues de l'IA ou « Big Data » (Cointet & Parasie, 2018) destinées à traiter de très grands corpus de données est apparu aux différents acteurs[3] comme la seule solution possible pour répondre à la nécessité d'agréger, de hiérarchiser et de classer un très grand nombre de propositions tout en respectant les réquisits de la démocratie participative et/ou délibérative. La conception même du type de démocratie engagée par ces débats et leurs implémentations numériques apparaît en effet comme un enjeu de recherche.

---

[1] https://granddebat.fr/

[2] https://le-vrai-debat.fr/

[3] https://www.lemonde.fr/pixels/article/2019/02/01/grand-debat-en-ligne-et-democratie-l-analyse-et-la-transparence-des-donnees-en-question_5417911_4408996.html

Les attentes à l'égard de ces approches apparaissent très importantes, peut-être disproportionnées compte-tenu de ce qu'elles sont réellement capables de faire (Boelaert & Ollion, 2018 ; Bellet et *al*., 2020) et relèveraient ainsi assez classiquement d'une sociologie de la promesse (Joly, 2015 ; Vinck, 2015). Outre l'enjeu politique d'une mise en discussion de ce que les algorithmes font à la démocratie (Cardon, 2015), il y a bien un enjeu scientifique à tenter de cerner ce que peuvent faire (et ce que ne peuvent pas faire) ces méthodes d'analyse automatique. La comparaison du Grand Débat National et du Vrai Débat présente ainsi un intérêt tant du point de vue de la sociologie politique que du point de vue de l'analyse des textes – que l'on considère les approches de textométrie ou de TAL.

Il existe d'ores et déjà une très abondante littérature en sciences sociales sur le mouvement des Gilets Jaunes et de nombreux travaux sont en cours[4]. De nombreuses questions portent sur la sociologie des participants à ces débats : les premiers résultats des travaux en cours suggèrent une sociologie très différente de ces deux publics. L'enquête du laboratoire PACTE[5] (Guerra *et al.* 2019 ; Abrial *et al.* 2020) réalisée auprès de groupes de Gilets Jaunes sur Facebook décrit un public majoritairement constitué de travailleurs précaires[6], habitant en territoire rural ou périurbain et refusant majoritairement de se situer sur l'axe gauche-droite. L'enquête CEVIPOF menée sur la population des participants aux réunions (RIL) du Grand Débat National dresse un portrait assez éloigné de ce premier profil : la moyenne d'âge des personnes interrogées est en effet de 57 ans, la moitié sont des retraités, 62% d'entre eux disent posséder un diplôme de l'enseignement supérieur, enfin 54% déclarent s'en « sortir plutôt facilement avec les revenus du ménage ». On note par ailleurs une surreprésentation des personnes habitant les grandes villes – notamment celles ayant placé en tête Macron lors du premier tour de l'élection présidentielle (CEVIPOF, 2019).

Ces enquêtes donnent des indications précieuses mais seulement indirectes : rien ne nous permet en effet d'extrapoler à partir de l'enquête sur les RIL ou sur les groupes Facebook GJ aux publics des débats numériques. Nous ne disposons pas de données directes décrivant le profil des participants aux deux plateformes de débat, la seule information disponible est le code postal pour le GDN.

Nous avons travaillé sur les corpus consacrés à la transition environnementale extraites de ces deux plateformes et sur un troisième corpus issus de la plateforme « Entendre la France ». Celle-ci a récolté par l'intermédiaire de Facebook les réponses au même questionnaire que celui proposé par le GDN. Il s'agit d'un public qui est probablement nettement plus jeune que celui du GDN – c'est d'ailleurs l'objectif poursuivi par les promoteurs de la plateforme.

---

[4] Les deux journées d'étude des 16 et 17 janvier 2020, « Quels outils pour l'analyse des Gilets Jaunes », attestent sous l'angle méthodologie de la diversité et inventivité des approches dans les travaux en cours.

[5] https://www.pacte-grenoble.fr/programmes/grande-enquete-sur-le-mouvement-des-gilets-jaunes

[6] 67% peuvent être considérés comme étant dans une « situation précaire » , le double de la moyenne nationale.

| *Taille* | *Vrai Débat*[7] *sans arguments* | *Vrai Débat et arguments* | *Grand Débat National* | *Entendre la France* |
|---|---|---|---|---|
| Nombre de textes | 2 599 | 6 373 | 87 552 | 39 430 |
| Nombre de formes | 17 707 | 22 380 | 78 829 | 34 582 |
| Nombre d'occurrences | 225 039 | 351 991 | 21 764 365 | 1 273 520 |
| Nombre de lemmes | 12 059 | 15 034 | 45 267 | 22 376 |

*Tableau 1 : Taille des corpus consacrés au thème de la transition environnementale*

A la différence du GDN, Entendre la France posait une série de questions socio-démographiques (notamment le sexe, l'âge) mais aussi celle du soutien aux Gilets Jaunes. La moitié des participants à la plateforme environ ont répondu à celles-ci. Elles constitueront des informations précieuses pour tenter de reconstituer les profils sociopolitiques des participants au Grand Débat National.

## 3. Un discours Gilets Jaunes au cœur du Grand Débat National ?

Notre premier objectif consiste à comparer les discours recueillis sur les plateformes GDN et VD à propos de la transition environnementale. Cette comparaison aura toutefois une portée limitée, elle portera sur les thématiques, les lexiques et donnera quelques aperçus sur les formes du discours. Elle se contentera de chercher des indices d'un discours GJ dans le GDN. Nous n'analyserons ni le détail des propositions (Zancarini et Ventresque, 2020), ni ne comparerons les énoncés par thématique (Loubère et Marchand, 2020). D'un point de vue méthodologique, nous adopterons une approche textométrique (spécificités puis classification hiérarchique descendante sous Iramuteq).

### *3.1 Analyse des spécificités*

L'analyse des spécificités fait apparaître une sur-représentation dans le GDN[8] des termes se rapportant au *dérèglement climatique* et de ses effets (*transition, sécheresse, érosion, déforestation*) au niveau *mondial* (*global, planète*). D'autres enjeux environnementaux sont aussi particulièrement évoqués comme la *biodiversité* et les causes de ces dérèglements comme la *pollution* (*pollueur, fossile, diesel, charbon* mais aussi les *pesticides*). On notera toutefois qu'une partie de ces problèmes ont été l'objet d'une injection thématique via la question fermée 1 (qui propose comme enjeux le dérèglement climatique (crue, sécheresse), l'érosion du littoral, la pollution de l'air, et la biodiversité et la disparition de certaines espèces). L'analyse des spécificités montre par ailleurs une surreprésentation de certains verbes comme *changer, réduire, diminuer* voire *arrêter* ou *interdire*. Elle met aussi en évidence un lexique propre au registre moral comme *prise de conscience, éduquer, éducation, responsable, habitude, effort, courage*… Les citoyens étant invités dans nombre de propositions à une prise de conscience des problèmes et à changer de comportement. Des verbes de modalisation comme *croire* ou *falloir* relèvent de ce registre déontique – on notera qu'il s'agit d'un registre introduit par le libellé de la question 2 « Que *faudrait*-il faire… ». Enfin, on note la surreprésentation d'un lexique macro-*économique* qui développe notamment le thème de la *croissance* (*investir, entreprise…*).

---

[7] Corpus disponible sur https://www.le-vrai-debat.fr/syntheses/

[8] Par ailleurs, une analyse des spécificités comparant l'ensemble des thématiques du GDN et du VD montre l'importance du lexique de la transition environnementale pour le GDN, le VD se caractérisant davantage par les thématiques portant notamment sur la réforme de la démocratie (referendum, constitution, etc.).

L'analyse des spécificités du VD fait apparaître une liste d'enjeux sensiblement différents : la question de la *vitesse* et des *radars* sur les *routes* et les *autoroutes* constitue ainsi un thème qui assez logiquement, compte-tenu de l'histoire du mouvement des Gilets Jaunes caractérise ce corpus[9]. Les questions liées à l'alimentation sont aussi caractéristiques de ces discours (*alimentaire, fruit, fruitier, légume, étiquetage, label paysan, producteurs, distributeurs, PAC*). Il est vrai que le libellé du thème sur la plateforme VD faisait explicitement allusion aux questions agricoles. La *souffrance animale* notamment dans les *abattoirs* (ainsi que les conditions de leur mise à *mort*) donne lieu à un lexique qui s'avère caractéristique du Vrai Débat par rapport au Grand Débat National. Les conditions de vie notamment économiques (*prix, euro, montant, facture, gratuité, vendre, vente*), la dénonciation d'un *marché* et de la *spéculation* sont également repérables à travers l'analyse des spécificités. Ce thème fait probablement écho à un discours sur des privatisations jugées indues (*privé, nationaliser, public, coopératif*) dont on sait qu'elles ont constitué un des thèmes de propositions de la plateforme (dénonciation de la privatisation d'Aéroport de Paris etc.). Le lexique se rapportant à la situation de la *sncf* (*ligne, réseau*) peut sans doute être rapproché de cette thématique. Enfin, on note une série de termes se référant à des controverses impliquant *EDF* notamment le *compteur Linky* mais aussi la question de l'enfouissement des déchets à *Bure*. Une série de verbes caractérise ces discours (*refuser, suspendre exiger* mais aussi *permettre*) dont un verbe de modalisation particulièrement employé : *pouvoir*.

### 3.2 Classification hiérarchique descendante

Une classification hiérarchique descendante sur l'ensemble du corpus formé par le GDN et le VD permet d'éclairer les thématiques privilégiées par l'une ou l'autre plateforme de débat. Toutefois compte-tenu de la différence de taille entre les deux sous-corpus (le corpus du GDN consacré aux thématiques environnementales est environ 60 fois plus important que celui du VD[10]), la structure thématique mise en évidence est clairement celle du GDN.

La première coupure de la classification permet de vérifier que c'est bien la structure thématique du GDN qui apparaît dans cette analyse : cette première coupure oppose en effet une vision macroscopique de la transition environnementale à une vision plus microscopique, centrée notamment sur les pratiques. Les questions du GDN invitent le répondant à se focaliser successivement sur ces deux types de visions, invitant à développer un discours en « je », centré sur les changements de comportements individuels (notamment question 38 à 42[11], puis à

---

[9] La limitation de la vitesse à 80km/h sur les routes (sans séparation entre les voies) constitue un des éléments déclencheurs de ce mouvement avec la question de la taxe *carbone*. Toutefois, cela ne signifie pas que les propositions de ce mouvement portent exclusivement, ni même prioritairement sur cet enjeu comme l'ont montré notamment Zancarini et Ventresque (2020).

[10] Les problèmes posés par la comparaison de ces deux corpus de tailles très différentes sont abordés dans Loubère et Marchand, (2020).

[11] Par exemple Q39. « *Quelles seraient pour vous les solutions les plus simples et les plus supportables sur un plan financier pour vous inciter à changer vos comportements ?* »

envisager les problèmes sous un angle plus collectif et à produire ainsi un discours en « nous » (33 et 34, puis 43 à 48)[12].

Cette branche de la classification plus « politique » se scinde d'une part, en un groupe de deux classes d'énoncés focalisés sur la biodiversité et l'agriculture et les produits bio et, d'autre part, en un groupe de cinq classes qui se scinde lui-même en deux classes très proches sur le changement climatique et l'énergie, et trois classes à tonalité soit universaliste (*monde, planète, humanité* etc.), soit discutant les conditions politiques – notamment au niveau européen – ou économiques de la transition environnementale. Toutes ces classes à l'exception de la classe sur le changement climatique s'avèrent caractéristiques du VD, *i.e.* sont caractérisées par la variable illustrative (mot étoilé dans le vocabulaire Reinert /Iramuteq) VD.

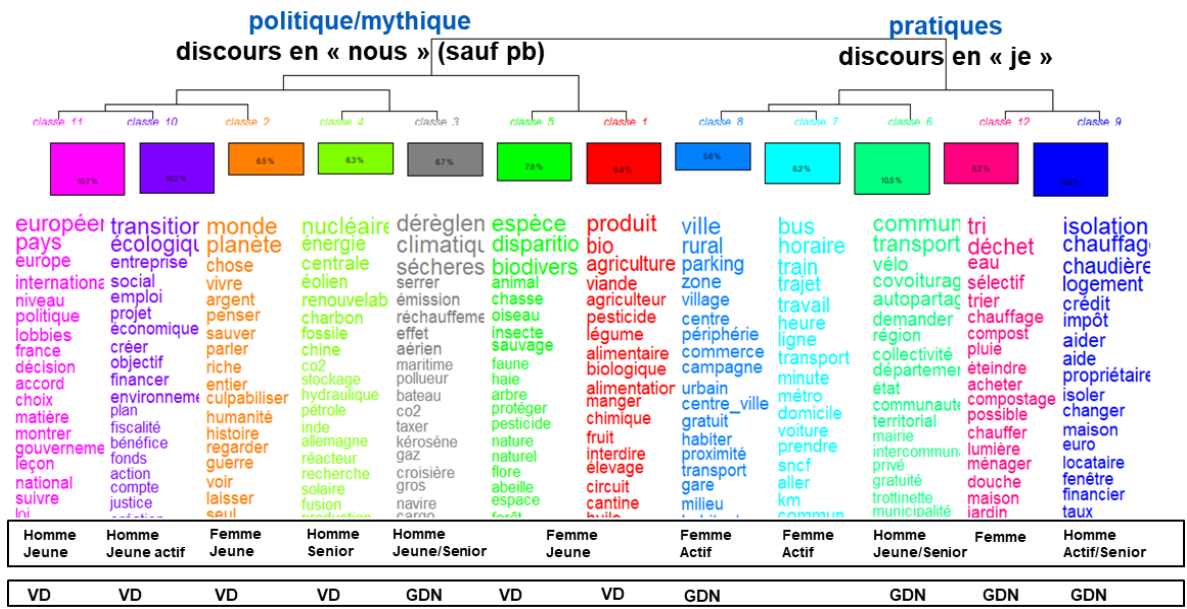

*Figure 1 : Dendrogramme de la classification descendante hiérarchique de l'ensemble des deux corpus de débat, pseudo variables et spécificités*

La branche de discours en « je » portant prioritairement sur un compte-rendu des pratiques réelles ou jugées souhaitables comprend deux sous-groupes d'énoncés : le premier composé de trois classes porte sur le transport (domicile-travail, rural-urbain et transport en commun en ville), le second porte sur des pratiques plus domestiques ou du moins liées au foyer puisqu'il s'agit d'une part de pratiques de tri des déchets ou de compost et d'autre part, d'isolation ou de chauffage. Toutes ces classes, à l'exception de celle développant la thématique du transport domicile/travail sont caractéristiques du GDN[13].

---

[12] Par exemple Q48. *« Que pourrait faire la France pour faire partager ses choix en matière d'environnement au niveau européen et international ? »*

[13] Cela ne signifie pas que la première personne du singulier soit plus utilisée dans GDN que dans le VD mais que celle-ci, dans le GDN, est assez systématiquement associée à un discours sur les pratiques personnelles, le « nous » étant réservé aux énoncés plus macroscopiques. On recense par ailleurs un emploi important de la première personne du singulier dans les énoncés d'arguments recueillis dans le VD. Ces énoncés sont en outre caractérisés par une surreprésentation de connecteurs de condition et de cause – analyse Tropes.

On constate ici d'une part, la congruence de ces résultats avec les premières indications données par l'analyse de spécificités, d'autre part, les effets de cadrage très importants des questions du GDN : en effet toutes les classes portant sur la pratique s'avèrent caractéristique du GDN. Le VD (le sous corpus est rappelons-le beaucoup plus petit) s'avère par contraste caractéristique des énoncés en « nous ».

## 4. Un public Gilets Jaunes au cœur du Grand Débat National ?

Pour caractériser les locuteurs, nous avons utilisé le corpus du site « Entendre la France »[14]. Il s'agit du site d'une association dont « *le but est de permettre au plus grand nombre de Français de s'exprimer de la manière la plus simple possible, et d'être entendus* ». Les utilisateurs pouvaient répondre aux questions du Grand Débat, directement sur le site Web mais également via Messenger. En plus des questions du Grand Débat, ils pouvaient renseigner les caractéristiques suivantes : code postal, commune, type de commune, sexe, âge, formation, profession, taille de l'organisation et position vis-à-vis des GJ.

Nous avons retenu les variables suivantes :
- Sexe : 2 catégories homme/femme ;
- Age : 7 catégories réorganisées en 4 : « jeune », « jeune actif », « actif » et « senior » ;
- Position vis-à-vis des GJ : 3 catégories réorganisées en 2 : « soutient/ne soutient pas ».

L'objectif était d'utiliser le contenu textuel des réponses pour prédire ces variables à l'aide de différentes techniques de « *Machine Learning* ».

### *4.1 Machine learning*

Pour prédire chacune des 3 variables, 3 couples « données d'apprentissage/données de test » ont été constitués[15], données réparties en 70% pour l'apprentissage et 30% pour le test. Les méthodes suivantes ont ensuite été testées :

*Bayésien naïf*

Il s'agit d'une méthode robuste et très rapide qui a fait ses preuves, par exemple dans le cadre de la détection des spams dans les mails. Basée sur le célèbre théorème de Bayes, la méthode calcule la probabilité d'apparition des mots dans les différentes catégories, exemple « soutient » et « soutient pas ». Pour prédire l'appartenance d'un texte à une catégorie, la méthode calcule la probabilité d'appartenance aux différentes catégories et l'affecte à celle dont la probabilité est la plus élevée.

*Régression logistique*

La régression logistique vise à prédire les valeurs d'une variable catégorielle binaire (homme/femme ou soutient/soutient_pas). Dans le cas de l'âge, pour lequel il y a 4 catégories, nous entrainons 4 classifieurs binaires de type *one against all*. Pour prédire l'appartenance d'un texte à une catégorie, les 4 classifieurs sont appliqués, le texte est affecté à la catégorie pour laquelle le score du classifieur est le plus élevé.

---

[14] https://www.entendrelafrance.fr/

[15] Les utilisateurs n'étaient pas tenus de renseigner tous les champs. Ont été gardées les réponses des personnes ayant renseigné leur âge pour prédire la variable « âge ».

*Word Embedding + Docov*

Dans un premier temps, un modèle Word2Vec (Mikolov *et al.*, 2013) est appris sur les données textuelles. Ensuite, les documents vont être transformés en vecteurs à l'aide de l'algorithme DoCoV pour « Document Co Variance » (Torki, 2018) qui calcule une représentation vectorielle du document à partir de la matrice de covariance des *embeddings* de mots. Ces méthodes ont montré qu'elles pouvaient fournir des représentations vectorielles de qualité (Suignard *et al.*, 2019).

Pour chaque catégorie (2 dans l'exemple du soutien aux Gilets Jaunes), est calculé le barycentre en moyennant les vecteurs des documents de chaque catégorie.

Pour prédire la catégorie d'un document, il suffit de calculer sa similarité avec le centre des différentes catégories et de l'affecter à la catégorie pour laquelle la similarité est la plus grande.

*Bert et CamemBERT*

Les méthodes précédentes ne tiennent pas compte de l'ordre des mots dans la phrase et c'est précisément ce qu'apportent les méthodes de type « *Masked Language Model* ». Nous avons utilisé Bert (Devlin *et al.*, 2018) (pour *Bidirectional Encoder Representations from Transformers*) et CamemBERT (Martin L *et al.*, 2019), une version de BERT entrainée sur des données françaises. Ces méthodes utilisent des architectures neuronales et sont entrainées sur des corpus volumineux et sur plusieurs tâches, une des tâches consistant à prédire un mot aléatoirement masqué. Après cette phase générique, une phase plus spécifique de « *transfert learning* » par « *fine-tuning* » est appliquée pour adapter plus finement les données au corpus cible.

Au final, ces méthodes permettent d'obtenir des représentations vectorielles des documents « au plus près des données » et sont utilisées dans différentes tâches comme la classification supervisée ou non-supervisée.

On constate que les 4 méthodes fournissent sensiblement les mêmes résultats. Néanmoins, BERT est un peu en dessous[16] et a tendance à affecter très majoritairement les documents dans la catégorie « Soutient », ce qui explique son faible score de rappel. Pour cette raison, seules les 3 premières méthodes sont conservées dans la suite.

|           | Naive Bayes | Regression | Docov | BERT  |
|-----------|-------------|------------|-------|-------|
| Précision | 0,664       | 0,648      | 0,645 | 0,680 |
| Rappel    | 0,667       | 0,645      | 0,648 | 0,609 |
| F-Mesure  | 0,665       | 0,646      | 0,641 | 0,589 |

*Tableau 2 : comparaison des 4 méthodes pour la prédiction du soutien aux GJ*

### *4.2 Enrichir l'analyse des données textuelles en documentant les propriétés des locuteurs*

Le corpus « Entendre la France » nous a donc permis d'entrainer des classifieurs à prédire l'âge des répondants, leur sexe et leur soutien ou non à la cause des Gilets Jaunes. Nous avons appliqué les classifieurs ainsi entrainés sur le corpus du Grand Débat, puis un système de vote nous a permis de conserver le vote majoritaire entre ces 3 classifieurs et de calculer 3 nouvelles variables étoilées ajoutées au fichier Iramuteq.

---

[16] Les mêmes scores sont obtenus avec CamemBERT. La faiblesse des résultats obtenus avec BERT et CamemBERT est un peu surprenante, étant donné que ces méthodes obtiennent de très bons résultats dans la plupart des tâches de classification.

Il est désormais possible d'attribuer ces propriétés socio-politiques reconstituées aux locuteurs du GDN. Bien sûr, cette attribution est hypothétique : elle suppose notamment de considérer que le public de la plateforme Entendre la France présente les mêmes propriétés sociolinguistiques que les locuteurs s'exprimant sur la plateforme du GDN. Or, nous avons toutes raisons de croire que ces deux publics diffèrent : la plateforme d'Entendre la France ayant été notamment créée pour pallier une participation supposée insuffisante des plus jeunes. Le profil des locuteurs d'Entendre la France est en effet beaucoup plus jeune que celui des participants aux réunions publiques tel que l'enquête CEVIPOF nous le restitue. La reconstitution de l'âge des participants du GDN double ainsi le nombre de jeunes actifs, (mais il divise par deux celui des seniors). Il est possible que le profil des seniors déclarés d'Entendre la France soit assez différent de celui des seniors du GDN. La prédiction de la variable de genre nous révèle par ailleurs un public du GDN sensiblement plus masculin que celui des participants à la plateforme Entendre la France. Enfin, l'algorithme nous donne une proportion de 36% de participants de la plateforme GDN soutenant le mouvement des GJ et de 64% ne le soutenant pas. Cette proportion apparaît vraisemblable, on s'attend en effet à une proportion forte de personnes ne soutenant pas les GJ dans le GDN lancé par le Président Macron et critiqué par ailleurs par les GJ. Toutefois, si cette proposition est minoritaire, elle est loin d'être négligeable – ce que nous laissait soupçonner une analyse même superficielle des discours issus du GDN. Cette proportion est par ailleurs comparable – environ 40% de soutien aux GJ – à celle trouvée par B. Monnery (2020) à partir d'une méthode différente (un modèle de régression construit à partir de l'analyse des réponses aux questions fermées sur Entendre la France pour expliquer le soutien aux Gilets Jaunes).

**Entendre la France**

| | | | |
|---|---|---|---|
| soutient | 56,3 | jeune | 62,3 |
| ne soutient pas | 43,7 | jeune actif | 19,3 |
| homme | 61,9 | actif | 10,5 |
| femme | 38,1 | senior | 7,9 |

**Prédiction appliquée au GDN**

| | | | |
|---|---|---|---|
| soutient | 36,0 | jeune | 39,0 |
| ne soutient pas | 64,0 | jeune actif | 40,8 |
| homme | 76,0 | actif | 16,5 |
| femme | 24,0 | senior | 3,7 |

*Tableaux 3 et 4 : composition sociopolitique du corpus « Entendre la France »[17] et de la prédiction appliquée au GDN - 87552 documents (en %)*

*4.3 Une prédiction interprétable ?*

L'approche que nous avons adoptée ici, pour mieux connaître les propriétés sociales des locuteurs du GDN, emprunte une voie probabiliste qui permet de qualifier directement le locuteur. Une autre approche « indirecte » mais certaine qui caractérise le contexte sociodémographique du locuteur à partir de son lieu d'habitation via les codes postaux est néanmoins possible.

L'approche par réseaux de neurones pose, on le sait, d'importants problèmes d'explicabilité, (notamment Bollaert et Ollion, 2018). A la différence des approches de type régression, il n'est pas possible de construire un modèle et de donner une valeur à différentes variables explicatives. En revanche, dans le cas qui nous occupe, il est possible de vérifier si de « pseudo variables », que l'on peut qualifier encore de « propriétés probabilistes » donnent lieu à des interprétations intéressantes du point de vue de la sociologie politique. La figure 1 présente la classification analysée plus haut mais caractérise aussi les classes d'énoncés avec les pseudo variables issus du machine learning (lignes en bas de la figure 1).

---

[17] % calculés sur le nombre de réponses qualifiées (âge : 10 637 ; sexe : 12 398 ; soutien : 6 278).

La plupart des classes d'énoncés « en nous », dont nous savons par ailleurs qu'elles sont caractéristiques du VD s'avèrent aussi caractéristiques des locuteurs qui soutiendraient les Gilets Jaunes selon l'algorithme d'apprentissage (à partir d'Entendre le France). Seule la classe sur le changement climatique, dont on sait par ailleurs qu'elle est caractéristique du GDN, serait le fait de locuteurs ne soutenant pas les Gilets Jaunes. Inversement toutes les classes d'énoncés portant sur les pratiques seraient caractéristiques de locuteurs ne soutenant pas les Gilets Jaunes. Ces résultats convergents avec les premières analyses tendent à conforter la qualité des pseudo-variables – au moins celle portant sur le soutien au mouvement des Gilets Jaunes.

L'analyse de la variable de genre et d'âge donne par ailleurs des résultats que l'on peut qualifier de « vraisemblables », ou de non contre-intuitifs : les femmes seraient ainsi surreprésentées parmi les locuteurs de la classe de discours universaliste (*monde/planète*), de la classe d'énoncés sur les produits bio et la biodiversité. Elles seraient aussi caractéristiques des classes transport rural-urbain et domicile-travail mais aussi tri des déchets. Les hommes seraient surreprésentés parmi les locuteurs des classes conditions politiques et économiques de la transition environnementale, énergie, transport en commun et isolation/chauffage. Le même type de commentaires pourrait être fait à propos de la pseudo-variable sur les classes d'âge.

## 5. Conclusions

A l'issue de ce parcours, plusieurs conclusions peuvent être tirées et sans doute autant de questions posées. Ces conclusions portent sur les discours, les effets du design des plateformes, la sociologie des participants et enfin les algorithmes et leurs usages.

Sur les discours, l'analyse met en évidence des cadrages différents des problèmes (GDN : dérèglement climatique, taxer les transports *vs* agriculture intensive/alimentation, croissance/décroissance/humanité/démographie, débats scientifiques/lobbys/politique pour le VD) mais aussi des registres plus politiques ou focalisés sur les comportements des individus, favorisant un registre plus moral ou privilégiant des mesures, des actions à entreprendre plutôt que la description des problèmes.

Toutefois, une partie importante de ces différences semble relever des effets de cadrage important du design des plateformes. Dans le GDN, des questions ouvertes injectent des thèmes et délimitent des espaces d'action collective et un espace d'actions individuelles, énoncés orientés problème/solution, quand le VD est caractérisé par un sous-ensemble d'énoncés destinés à soutenir ou combattre des propositions par rapport l'objet de vote. Il convient ici de souligner une limite importante de cette analyse qui ne prend pas en compte la dimension délibérative de cette plateforme visant notamment à organiser la hiérarchisation des propositions par l'échange d'arguments et le vote. L'analyse des discours telle qu'elle a été conduite ne reflète pas cette hiérarchisation, la hiérarchie des thèmes et des classes ne reflétant pas la hiérarchie des votes. Une voie possible consisterait à caractériser chacun des énoncés analysés dans le VD par le poids des votes, solution que nous devons à Loubère et Marchand (2020). Malgré ces effets de cadrage non négligeables, l'analyse fait apparaître dans le GDN les marques d'un discours proche des Gilets Jaunes dans la mesure où il reflète les thèmes et le discours analysés sur la plateforme VD.

Concernant la sociologie des publics, nous avons proposé une approche par *machine learning* qui permet d'inférer certaines des propriétés socio-politiques des locuteurs du GDN. Cette analyse sur le fond montre sinon la présence de locuteurs GJ, du moins de soutien à ce mouvement. La question de la sociologie des Gilets Jaunes et de leurs soutiens reste très largement ouverte, on sait par ailleurs qu'elle a probablement varié dans le temps, ce qui plaiderait pour une analyse diachronique du corpus.

La question de la reconstitution de données absentes (pseudo variables reconstituant les propriétés sociales et politiques des locuteurs) est aussi un chantier à peine exploré : la plus grande prudence est de mise quant aux conditions de ces inférences. Une des voies possibles nous semble, à défaut d'explicabilité des approches de *machine learning* de discuter de leur interprétabilité à la lumière notamment des travaux en sociologie politique sur le mouvement des Gilets Jaunes et sur le Grand Débat National.

Enfin, ces travaux posent une série de questions sur la comparaison entre des approches de type ADT et IA mais aussi leur combinaison. Sans les détailler ici, on soulignera les vertus des approches Reinert sous Iramuteq qui permettent de hiérarchiser des cadrages, celles plus générales de la textométrie qui propose de calculer le poids des variables lexicales ou des formes de discours (par exemple les marques de l'argumentation). D'autres recherches montrent que les algorithmes de plongement de mots s'avèrent par ailleurs des outils d'exploration à la fois puissants et fins des thématisations par l'identification de champs lexicaux (Brugidou et *al.*, 2019).

## 6. Références